\documentclass[twocolumn,showpacs,preprintnumbers,aps,prl,amsmath,amssymb,floatfix,superscriptaddress]{revtex4-1}

\usepackage{graphicx}
\usepackage{dcolumn}
\usepackage{bm}
\usepackage{epsfig}
\usepackage{color}
\usepackage{natbib}

\newcommand{\ket}[1]{\left|{#1}\right\rangle}
\def\Rb{$^{87}$Rb }
\def\Na{$^{23}$Na }
\def\NaRb{$^{23}$Na$^{87}$Rb }

\def\ket#1{\left|#1\right\rangle}
\def\Xstate{$X^1\Sigma^+$}

\begin{document}

\title{Creation of an ultracold gas of ground-state dipolar $^{23}$Na$^{87}$Rb molecules}

\author{Mingyang Guo}
\affiliation{Department of Physics, The Chinese University of Hong Kong, Hong Kong, China}
\author{Bing Zhu}
\affiliation{Department of Physics, The Chinese University of Hong Kong, Hong Kong, China}
\author{Bo Lu}
\affiliation{Department of Physics, The Chinese University of Hong Kong, Hong Kong, China}
\author{Xin Ye}
\affiliation{Department of Physics, The Chinese University of Hong Kong, Hong Kong, China}
\author{Fudong Wang}
\affiliation{Department of Physics, The Chinese University of Hong Kong, Hong Kong, China}
\author{Romain Vexiau}
\affiliation{Laboratoire Aim\'e Cotton, CNRS, Universit\'e Paris-Sud, ENS Cachan, Universit\'e Paris-Saclay, 91405 Orsay Cedex, France}
\author{Nadia Bouloufa-Maafa}
\affiliation{Laboratoire Aim\'e Cotton, CNRS, Universit\'e Paris-Sud, ENS Cachan, Universit\'e Paris-Saclay, 91405 Orsay Cedex, France}
\author{Goulven Qu\'em\'ener}
\affiliation{Laboratoire Aim\'e Cotton, CNRS, Universit\'e Paris-Sud, ENS Cachan, Universit\'e Paris-Saclay, 91405 Orsay Cedex, France}
\author{Olivier Dulieu}
\affiliation{Laboratoire Aim\'e Cotton, CNRS, Universit\'e Paris-Sud, ENS Cachan, Universit\'e Paris-Saclay, 91405 Orsay Cedex, France}
\author{Dajun Wang}
\email{djwang@phy.cuhk.edu.hk}
\affiliation{Department of Physics, The Chinese University of Hong Kong, Hong Kong, China}
\affiliation{The Chinese University of Hong Kong Shenzhen Research Institute, Shenzhen, China} 

\date{\today}
             
\begin{abstract}

We report the successful production of an ultracold sample of absolute ground-state \NaRb molecules. Starting from weakly-bound Feshbach molecules formed via magneto-association, the lowest rovibrational and hyperfine level of the electronic ground state is populated following a high efficiency and high resolution two-photon Raman process. The high purity absolute ground-state samples have up to 8000 molecules and densities of over $10^{11}$ cm$^{-3}$. By measuring the Stark shifts induced by external electric fields, we determined the permanent electric dipole moment of the absolute ground-state \NaRb and demonstrated the capability of inducing an effective dipole moment over one Debye. Bimolecular reaction between ground-state \NaRb molecules is endothermic, but we still observed a rather fast decay of the molecular sample. Our results pave the way toward investigation of ultracold molecular collisions in a fully controlled manner, and possibly to quantum gases of ultracold bosonic molecules with strong dipolar interactions.

    
\end{abstract}


\maketitle

Ultracold polar molecules (UPMs) featuring large electric dipole moments are promising candidates to investigate the physics of dipolar quantum gases~\cite{Trefzger2011,Baranov12}. However, the production of ground-state UPM is challenging due to the lack of efficient direct laser cooling of molecules. Currently, the most successful way of creating UPM is via magneto-association of ultracold atoms with the help of a Feshbach resonance~\cite{Kohler2006,Chin10}. With a properly designed stimulated Raman adiabatic passage (STIRAP)~\cite{Bergman98}, weakly bound Feshbach molecules formed this way can be transferred with high efficiency to the ground state where these molecules are more stable and the permanent electric dipole moment is large. This scheme was successfully demonstrated with fermionic $^{40}$K$^{87}$Rb~\cite{Ni2008} and $^{23}$Na$^{40}$K~\cite{Park2015} molecules, and with bosonic $^{87}$Rb$^{133}$Cs molecules~\cite{Takekoshi2014,Molony2014}.

Bialkali UPMs can be divided into two classes depending on their chemical stability~\cite{Zuchowski2010} in their absolute ground state level. Two-body chemical reaction can happen in $^{40}$K$^{87}$Rb~\cite{Ospelkaus10,Ni10}, while it is energetically forbidden in $^{87}$Rb$^{133}$Cs and $^{23}$Na$^{40}$K~\cite{Zuchowski2010}. Signature of enhanced stability is actually observed in fermionic $^{23}$Na$^{40}$K~\cite{Park2015}, but not in bosonic $^{87}$Rb$^{133}$Cs~\cite{Takekoshi2014}. Currently, for both classes of molecules, the existence of other loss mechanisms, such as the enhanced loss from long-lived binary complexes~\cite{Mayle2013}, cannot be ruled out~\cite{Takekoshi2014,Park2015}. 

In this manuscript, we report the successful production of absolute ground-state \NaRb molecules by STIRAP. This bosonic molecule is also chemically stable in its rovibrational ground state \Xstate$\ket{v''=0, J''= 0}$, with the reaction 2$^{23}$Na$^{87}$Rb$\rightarrow$ $^{23}$Na$_2$+$^{87}$Rb$_2$ endothermic by 47 cm$^{-1}$~\cite{Zuchowski2010,Pashov2005,Seto2000,Jones96,note0}. Besides, it has a large permanent electric dipole moment~\cite{Dagdigian1972,IgelMann1986,Tarnovsky1993,Aymar2005}, which can lead to strong dipolar interactions. We achieved a full control of the population distribution among the nuclear spin hyperfine states, allowing us to create molecules in the absolute ground state with high purity. With two-photon Stark spectroscopy, we have measured the permanent electric dipole moment and demonstrated an induced dipole moment of $\mu$ = 1.06 D, higher than all previous works~\cite{Ni2008,Takekoshi2014,Molony2014,Park2015}. With this dipole moment, the dipolar interaction between two \NaRb molecules has a dipole length $l_D=M \mu^2/4\pi\epsilon_0\hbar^2$~\cite{Gao2008} as large as 35000 $a_0$. Here $M$ is the mass of the molecule, $\hbar$ is Planck constant divided by 2$\pi$, and $a_0$ is the Bohr radius.     

Same as our Feshbach molecule creation work~\cite{Wangfudong2015}, the current experiment starts from an ultracold mixture of \Na and \Rb atoms in their $\ket{F = 1, m_F=1}$ hyperfine states prepared in a 1064 nm optical trap. Here $F$ and $m_F$ are the quantum numbers of the atomic hyperfine state and its projection onto the magnetic field, respectively. Feshbach molecules are created by sweeping the magnetic field down to 335.6 G across the inter-species Feshbach resonance located at 347.7 G~\cite{Wangfudong2013}. For detection, we reverse the magnetic field sweep to dissociate the Feshbach molecules and absorption image the resulting atoms. All data in this work are taken by imaging Rb, because of the better signal-to-noise ratio.

By upgrading the setup, we significantly increased the number of Feshbach molecules from $2\times 10^3$~\cite{Wangfudong2015} to $10^4$. These molecules are mainly of triplet character resulting from the closed channel quantum state $a^3\Sigma^+\ket{v=21, J=1}$ with a dominating closed channel fraction and a binding energy of $\sim 2\pi\times 21$ MHz. Importantly, they also have a near zero magnetic dipole moment, which allows us to obtain a very pure molecular sample by removing residual atoms with a strong magnetic field gradient~\cite{Wangfudong2015}. This character is also advantageous for maintaining the two-photon resonance necessary for stable STIRAP, as molecules in the \Xstate state also have a near zero magnetic moment. Thus small magnetic field noises can only change the one-photon detuning by shifting the intermediate level, which affects the STIRAP efficiency much less sensitively.  

\begin{figure}[hbtp]
\includegraphics[width=0.85 \linewidth]{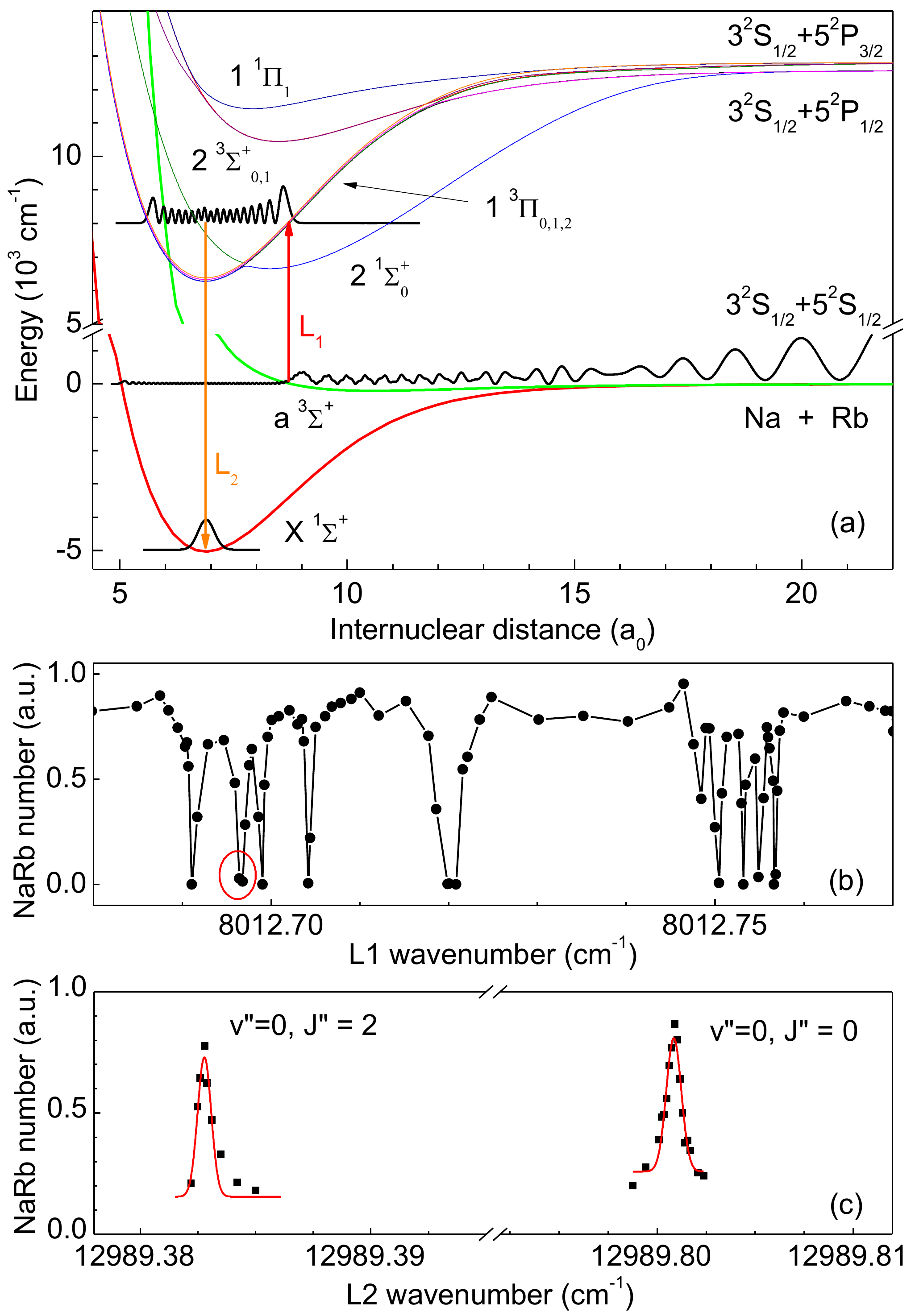}
\caption{\label{fig1}(color online). (a) Scheme of the NaRb potential energy curves including spin-orbit coupling, with squared density probabilities of the relevant levels. We used experimental potentials from Ref.~\cite{Pashov2005,Wangfudong2013} for the $X^1\Sigma^+$ and $a^3\Sigma^+$ states, and from Ref.~\cite{Docenko2007} for the 2$^1\Sigma^+$/1$^3\Pi_{\Omega=0,1,2}$ coupled states. The two-photon path through the 2$^1\Sigma^+$/1$^3\Pi$ states driven by the pump ($L_1$) and dump ($L_2$) lasers is shown schematically. (b) One-photon spectrum of the 1$^3\Pi$/2$^1\Sigma^+\ket{v' = 55, J' = 1}$ level near 8012.7 cm$^{-1}$, taken by scanning a free running pump laser. The selected intermediate level is marked by the red circle. (c) Two-photon spectroscopy for finding the lowest vibrational level $v'' = 0$ of the \Xstate state. Only the $J'' = 0$ and 2 rotational levels can be accessed because of parity selection rules.}
\end{figure}

For efficient population transfer, we built a Raman laser system consisting of two stabilized external-cavity diode lasers. With the standard Pound-Drever-Hall technique~\cite{Drever1983}, they are locked simultaneously to a 10 cm long and dual-wavelength coated ultrastable optical cavity with measured finesses of more than 25000 for both wavelengths. To our best estimate, the linewidths for both lasers are about 5 kHz. Both lasers are linearly polarized and co-propagate perpendicularly to the magnetic field. Each laser can thus drive $\pi$ or $\sigma$ (both $\sigma^+$ and $\sigma^-$) transition by setting its polarization parallel or perpendicular to the magnetic field.


To transfer the triplet Feshbach molecules to the singlet $X^1\Sigma^+$ state, the intermediate level has to be strongly triplet and singlet mixed~\cite{Ni2008,Takekoshi2014,Molony2014,Park2015}. As illustrated in Fig.~\ref{fig1}(a), for NaRb, such mixing can be found readily in the 1$^1\Pi/2^3\Sigma^+$~\cite{Kasahara1996} or the 2$^1\Sigma^+$/1$^3\Pi$~\cite{Docenko2007} coupled electronic states. Based on available experimental potentials of the 2$^1\Sigma^+$/1$^3\Pi$ states~\cite{Docenko2007}, we have identified several triplet and singlet mixed vibrational levels with large enough transition dipole moments to both the Feshbach state and the ground state. In Fig.\ref{fig1}(b), we show the one-photon spectrum of the transition from the Feshbach state to the coupled 2$^1\Sigma^+$/1$^3\Pi\ket{v' = 55, J' = 1}$ level near 8012.7 cm$^{-1}$~\cite{note1}. This level has $95\%$ 1$^3\Pi_0$ and $5\%$ 2$^1\Sigma^+$ character, and its hyperfine structures can be resolved and modeled~\cite{note1}. The spectrum shown is taken with the pump laser polarization aligned 45$^{\circ}$ to the magnetic field, thus both $\pi$ and $\sigma$ transitions can be accessed. One of the strongest $\pi$ transitions is selected as the intermediate level for STIRAP. The lifetime of this level is calibrated to be 238(30) ns~\cite{note1}, much longer than the $\sim$20 ns lifetime of typical excited levels. Fortunately, the measured transition dipole moment is $7.4(6)\times 10^{-4}~ea_0$, still large enough to support sub-MHz Rabi frequencies necessary for an efficient STIRAP in the current configuration.


The small 2$^1\Sigma^+$ component in the $v'$ = 55 level makes transition to the \Xstate state allowed. As shown in Fig.~\ref{fig1}(c), scanning a strong dump laser with the pump laser locked on resonance, we observed both the $J''$ = 0 and $J''$ = 2 rotational levels manifested as population gain caused by the Autler-Townes effect. The rotational constant $B_v$ obtained from this measurement is $2\pi\times 2.0896(5)$ GHz. We also found $v'' = 1$ at 105.833(3) cm$^{-1}$ above the $v'' = 0$ level. Both numbers are in excellent agreement with known values~\cite{Wangfudong2013}. The dissociation energy of the $v''$ = 0 level, directly measured from the energy difference between the two Raman lasers, is $D_0^X = 4977.308(3)$ cm$^{-1}$ (at zero magnetic field with respect to the hyperfine center of gravity), 0.018 cm$^{-1}$ less than the previously reported value~\cite{Wangfudong2013}. The quoted uncertainty of $D_0^X$ is mainly from the absolute accuracy of our wavelength meter~\cite{Zhu2016}.

\begin{figure}[hbtp]
\includegraphics[width=0.85 \linewidth]{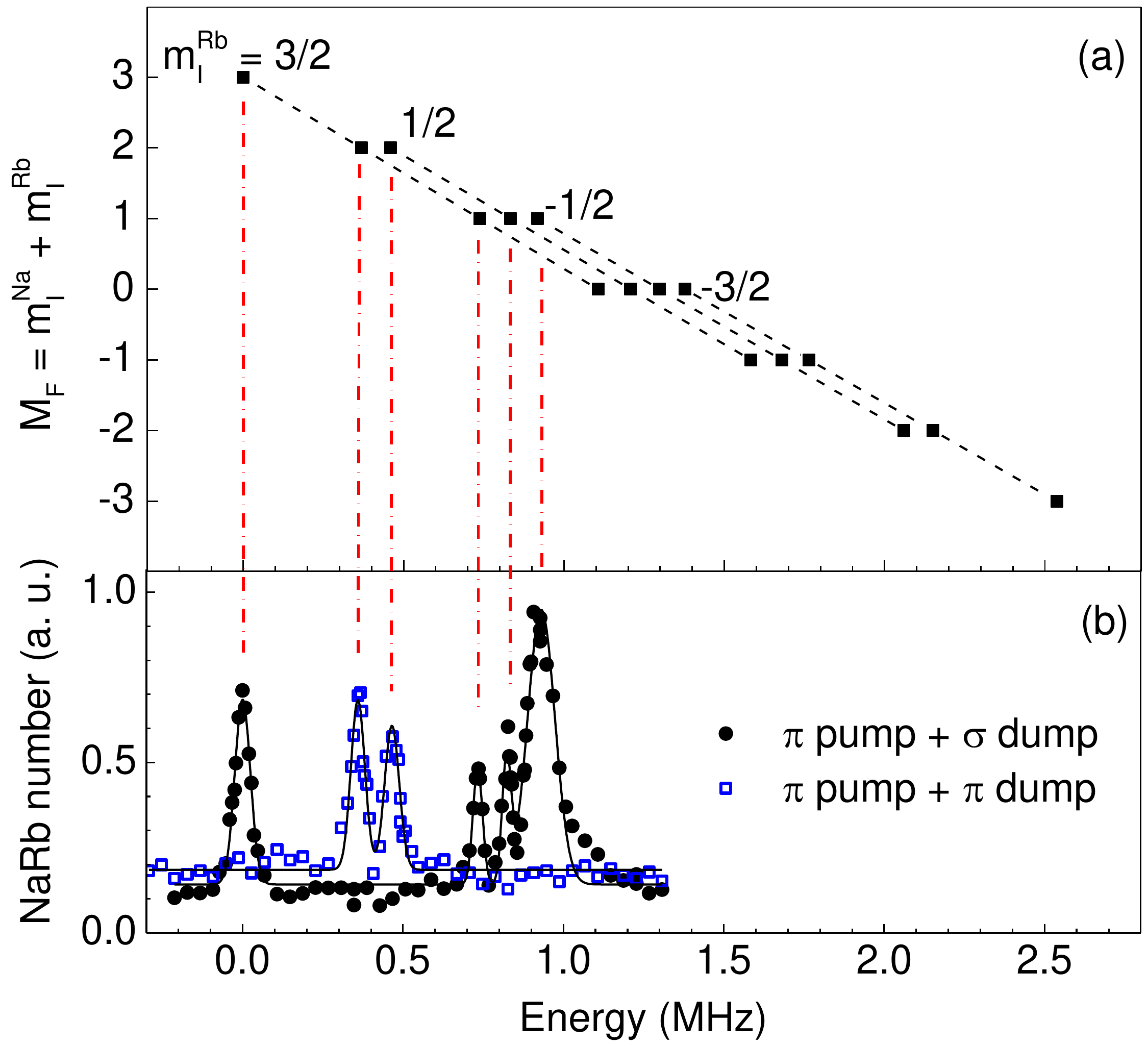}
\caption{\label{fig2}(color online). 
 (a) The calculated hyperfine Zeeman structures of the lowest rovibrational level of \NaRb at 335.6 G. The dashed lines indicate levels with the same $m_I^{Rb}$. (b) Two-photon spectrum obtained by dark resonance spectroscopy with 6 of the 16 hyperfine levels fully resolved. For this measurement, both lasers are locked to the optical cavity and a low dump transition Rabi frequency of $2\pi \times 170 $ kHz is used. Solid lines are from Gaussian fitting to the data points.}
\end{figure}

The hyperfine structure of the $X^1\Sigma^+\ket{v'' = 0, J'' = 0}$ state of bialkali molecule has been modeled in~\cite{Aldegunde2009,note3}. As shown in Fig.~\ref{fig2}(a), since both \Na and \Rb have nuclear spins of $I = 3/2$, there are 16 hyperfine Zeeman levels grouped according to the total angular momentum projection $M_F=m_I^{\rm{Na}}+m_I^{\rm{Rb}}$. As the Feshbach molecule has $M_F = 2$,  the absolute ground state with $M_F=3$ can be reached by the two-photon ($\pi$, then $\sigma$) transition directly. As shown in Fig.~\ref{fig2}(b), we resolved both the $M_F=3$ and the three $M_F=1$ levels unambiguously. By changing the polarizations to both $\pi$, the $M_F = 2$ hyperfine levels can also be fully resolved. The observed spacings between these observed hyperfine levels agree with the theoretical calculations almost exactly~\cite{Aldegunde2009,note3}, as depicted by the vertical dash-dotted lines. From the dark resonance lineshape, the transition dipole moment to the absolute ground state is determined to be $2.1(1)\times 10^{-2}~ea_0$~\cite{note1}, much stronger than the pump one.

\begin{figure}[hbtp]
\includegraphics[width=0.85 \linewidth]{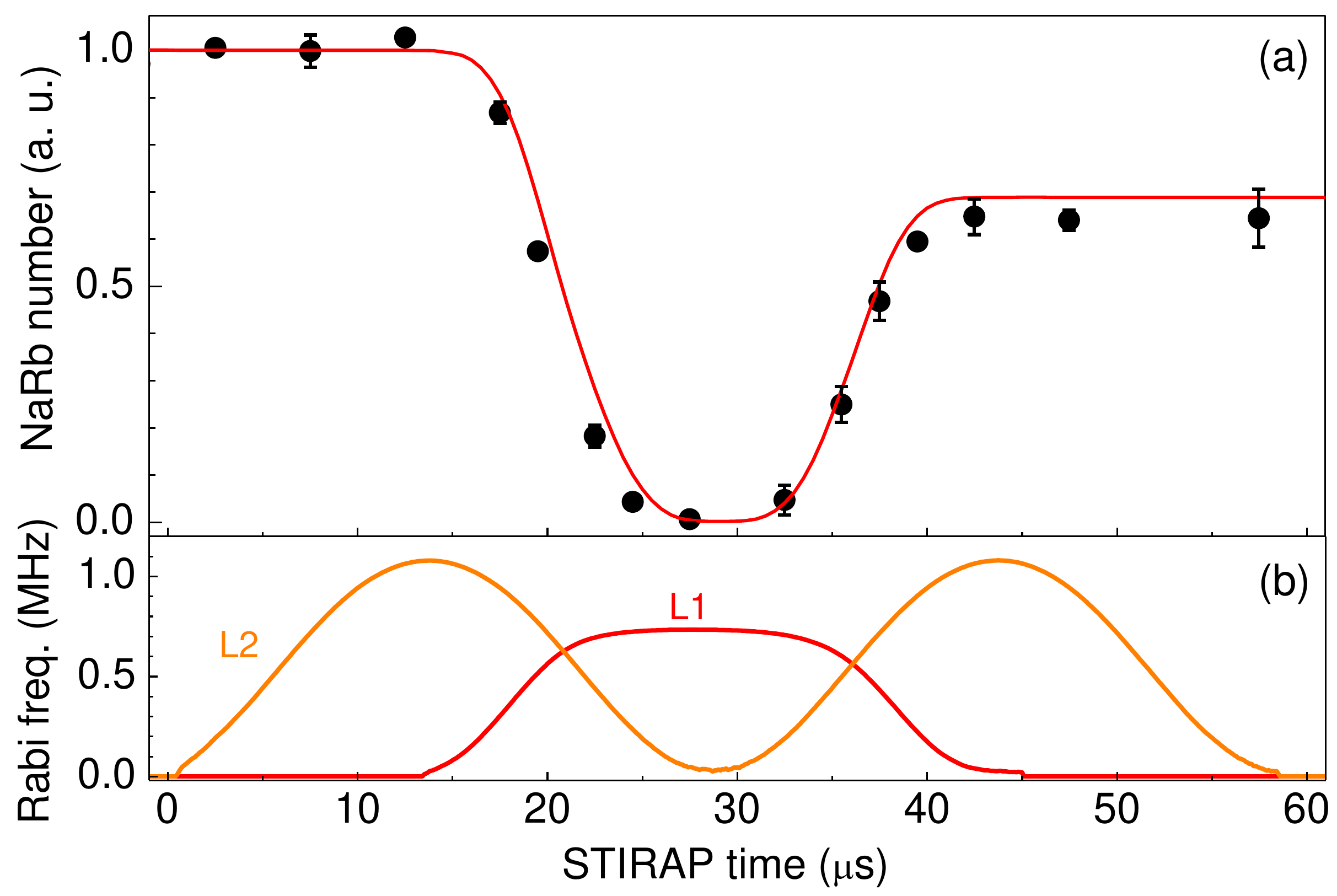}
\caption{\label{fig3}(color online). (a) Time evolution of the number of Feshbach molecules during a round-trip STIRAP. The data shown are taken with a SIIRAP pulse duration of 15 $\mu$s and a Rabi frequency of $2\pi\times 0.7(1.1)$ MHz for the pump(dump) transition. The reversed STIRAP is necessary for detection. The solid curve is the calculation using the master equation with experimentally measured parameters. (b) Pulse sequences for a round-trip STIRAP. The Rabi frequencies are from real measurements.  }
\end{figure}

To populate the absolute ground state, we perform STIRAP with the Raman laser pulse sequence shown in Fig.~\ref{fig3}(b) with both lasers locked on the corresponding resonances. A time-dependent dark state, which starts as the Feshbach state and ends as the absolute ground state, is created in the coupled three-level system during this sequence. After this sequence, the Feshbach molecules are transferred to the ground state adiabatically, as evident by the loss of signals at around 20 $\mu$s in Fig.~\ref{fig3}(a). As we cannot detect the ground-state molecules directly, a reversed STIRAP sequence is applied subsequently to bring them back to Feshbach molecules for detection. Assuming identical efficiencies for the two STIRAP processes, a one-way transfer efficiency of 80$\%$ can be inferred from the round-trip efficiency of about 64$\%$. As shown in Fig.~\ref{fig3}(a), this performance matches our simulation using the master equation with measured experimental parameters and estimated laser linewidths well. The simulation also indicates that transfer efficiencies over 90$\%$ are attainable by increasing the Rabi frequencies which are currently limited by the pump laser power.


\begin{figure}[hbtp]
\includegraphics[width=0.85 \linewidth]{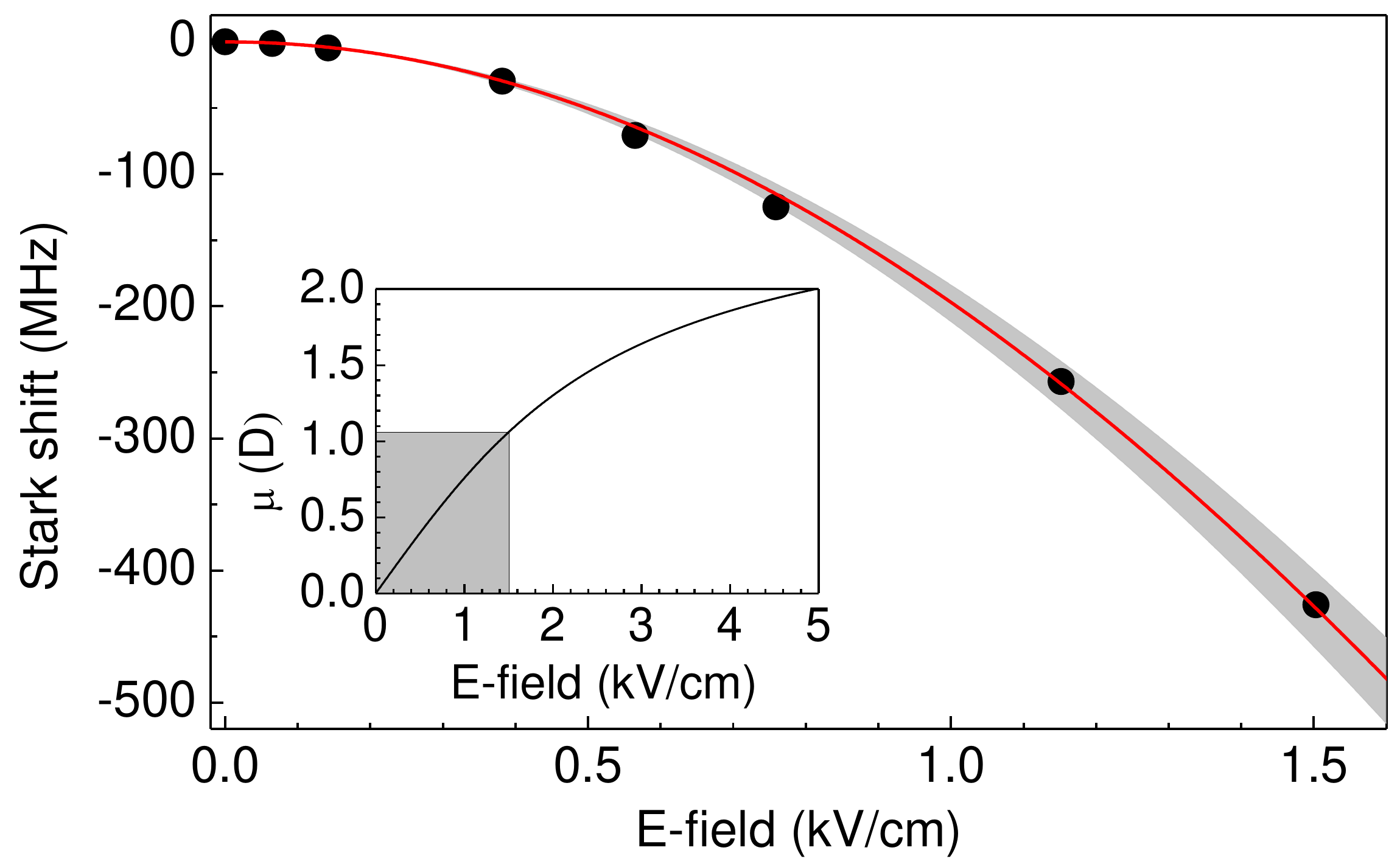}
\caption{\label{fig4}(color online). 
Stark shift of absolute ground-state \NaRb measured by dark resonance spectroscopy. Electric fields are calculated from the applied voltages and the effective spacing between the electrodes with all surrounding components taking into account. The solid curve is the fit to a model including contributions from several higher rotational levels. The uncertainty shown by the gray shading is mainly from the electric field determination. The inset shows the induced dipole moment vs the electric field with the currently accessible region marked by the shading area. }
\end{figure}

To illustrate the dipolar nature of \NaRb, we have installed two parallel-plate electrodes made of conductive and transparent ITO (Indium-Tin-Oxide) glasses outside of the vacuum cell. With these electrodes, an external electric field can be applied to induce an effective electric dipole moment in the laboratory frame. The Stark shift vs. electric field for the absolute ground state measured with dark resonance  is shown in Fig.~\ref{fig4}. This shift corresponds directly to the two-photon resonance shift, as the Feshbach molecule has essentially zero electric dipole moment. We also observed large shifts of the one-photon resonance from Stark effect of the common intermediate level. From fit (solid curve) of the absolute ground-state Stark shift to a model including contributions from rotational levels up to $J'' = 10$, we obtain a permanent electric dipole moment of $3.2(1)$ D, consistent with previously reported values~\cite{Dagdigian1972,IgelMann1986,Tarnovsky1993,Aymar2005}. A dipole moment of 1.06(4) D can already be induced in the current configuration (inset of Fig.~\ref{fig4}).

\begin{figure}[hbtp]
\includegraphics[width=0.85 \linewidth]{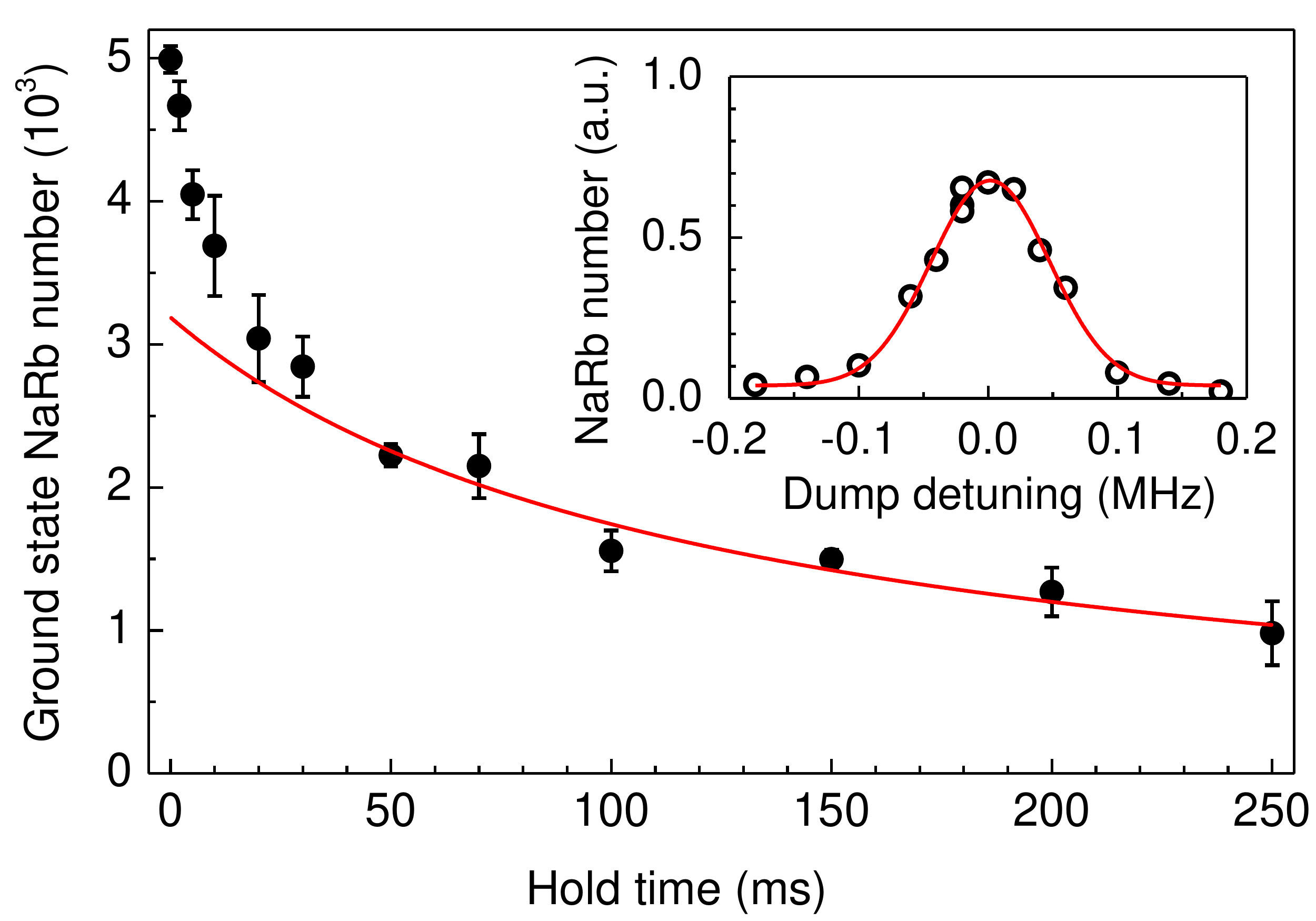}
\caption{\label{fig5}(color online). 
Decay of absolute ground-state \NaRb at zero electric field. The solid line is our fit based on a two-body decay model (see text). The measured trap frequencies for the ground-state molecules are $(\omega_x,\omega_y,\omega_z) = 2\pi\times (30,156,162)$ Hz~\cite{note2}. The sample has a typical temperature of 720 nK and an initial peak density of $1.5\times 10^{11}$ cm$^{-3}$. Error bars represent one standard deviation from three shots. The inset shows a typical round-trip STIRAP lineshape obtained by scanning the dump laser with the pump laser on resonance. }
\end{figure}


Despite their expected chemical stability~\cite{Zuchowski2010}, we observed a fast decay of the absolute ground state population by varying the holding time between the two STIRAP sequences (Fig.~\ref{fig5}). Fitting this decay to the two-body model $\dot{N} = -\beta (M \bar{\omega}^2/4\pi k_B T)^{3/2} N^2$, we extracted a loss rate coefficient $\beta = 2.5(9)\times 10^{-10}$ cm$^3$s$^{-1}$ (equivalent to a lifetime of about 0.2 s). Here we assumed a constant temperature $T$ and a Gaussian density profile, with $\bar{\omega}$ the mean trapping frequency, $k_B$ the Boltzmann constant. The uncertainly of $\beta$ is mainly from the number calibration. Only data after 50 ms are used in the two-body fit, as substantial breathing and sloshing motions can be clearly seen right after STIRAP~\cite{note2}. 

The origin of this large loss is not clear without knowing the final products. The typical linewidth of the STIRAP lineshape (the inset of Fig.~\ref{fig5}) is much smaller than the hyperfine level spacings, which ensures that all the transferred molecules are in a single and the lowest hyperfine level. However, due to the imperfect STIRAP, $\sim 20\%$ of the Feshbach molecules are not transferred to the ground state. No leftover molecules can be detected after the first STIRAP only, thus these molecules may distribute among a range of states and collide with the absolute ground-state molecules inelastically with an unknown rate. Another contribution could come from the so-called ``sticky mechanism'' which causes a unity short-range loss probability by forming long-lived complexes~\cite{Mayle2013}. Using the best known C$_6$ between two ground-state \NaRb molecules~\cite{Lepers2013}, the estimated rate~\cite{Wanggaoren2015} for this mechanism is $\beta^{sticky}_{GS-GS}(T=720~\rm{nK})=3.5\times 10^{-10}$ cm$^3$s$^{-1}$. Finally, the two-molecules photoassociation into an excited tetramer state~\cite{Lepers2013} by the trap light cannot be excluded neither, although this can only happen at short ranges since there are no resonances at long ranges.

This unexpectedly short lifetime imposes a challenge for many future works. More investigations are necessary to fully understand and possibly to mitigate this problem. For instance, we may use short-range shielding mechanisms such as $J''=1$ \NaRb molecules in a static electric field~\cite{Avdeenkov2006,Wanggaoren2015,Quemener2016} to suppress the losses at short distances. Ultimately, we can use optical lattices to isolate the molecules from each other to completely remove loss process~\cite{Danzl2010,Chotia2012}. With the current dipole moment, the nearest neighboring dipole-dipole interaction is already $> k_B\times 50$ nK in squared lattices formed by a 1064 nm laser. With such a strong long-range interaction, some of the predicted new quantum phases of the extended Bose-Hubbard model may already be observable~\cite{Buchler2007,Yi2007,Capogrosso2010}.

We thank J. Aldegunde and J. M. Hutson for providing us the hyperfine constants of the  NaRb molecule in the ground state. We are also grateful to K. Bergmann for the critical discussions on STIRAP. This work is supported by the COPOMOL project which is jointly funded by Hong Kong RGC (grant no. A-CUHK403/13) and France ANR (grant no. ANR-13-IS04-0004-01). The Hong Kong team is also supported by the RGC General Research Fund (grant no. CUHK404712) and the National Basic Research Program of China (grant No. 2014CB921403).


\begin{thebibliography}{39}%
\makeatletter
\providecommand \@ifxundefined [1]{%
 \@ifx{#1\undefined}
}%
\providecommand \@ifnum [1]{%
 \ifnum #1\expandafter \@firstoftwo
 \else \expandafter \@secondoftwo
 \fi
}%
\providecommand \@ifx [1]{%
 \ifx #1\expandafter \@firstoftwo
 \else \expandafter \@secondoftwo
 \fi
}%
\providecommand \natexlab [1]{#1}%
\providecommand \enquote  [1]{``#1''}%
\providecommand \bibnamefont  [1]{#1}%
\providecommand \bibfnamefont [1]{#1}%
\providecommand \citenamefont [1]{#1}%
\providecommand \href@noop [0]{\@secondoftwo}%
\providecommand \href [0]{\begingroup \@sanitize@url \@href}%
\providecommand \@href[1]{\@@startlink{#1}\@@href}%
\providecommand \@@href[1]{\endgroup#1\@@endlink}%
\providecommand \@sanitize@url [0]{\catcode `\\12\catcode `\$12\catcode
  `\&12\catcode `\#12\catcode `\^12\catcode `\_12\catcode `\%12\relax}%
\providecommand \@@startlink[1]{}%
\providecommand \@@endlink[0]{}%
\providecommand \url  [0]{\begingroup\@sanitize@url \@url }%
\providecommand \@url [1]{\endgroup\@href {#1}{\urlprefix }}%
\providecommand \urlprefix  [0]{URL }%
\providecommand \Eprint [0]{\href }%
\providecommand \doibase [0]{http://dx.doi.org/}%
\providecommand \selectlanguage [0]{\@gobble}%
\providecommand \bibinfo  [0]{\@secondoftwo}%
\providecommand \bibfield  [0]{\@secondoftwo}%
\providecommand \translation [1]{[#1]}%
\providecommand \BibitemOpen [0]{}%
\providecommand \bibitemStop [0]{}%
\providecommand \bibitemNoStop [0]{.\EOS\space}%
\providecommand \EOS [0]{\spacefactor3000\relax}%
\providecommand \BibitemShut  [1]{\csname bibitem#1\endcsname}%
\let\auto@bib@innerbib\@empty
\bibitem [{\citenamefont {Trefzger}\ \emph {et~al.}(2011)\citenamefont
  {Trefzger}, \citenamefont {Menotti}, \citenamefont {Capogrosso-Sansone},\
  and\ \citenamefont {Lewenstein}}]{Trefzger2011}%
  \BibitemOpen
  \bibfield  {author} {\bibinfo {author} {\bibfnamefont {C.}~\bibnamefont
  {Trefzger}}, \bibinfo {author} {\bibfnamefont {C.}~\bibnamefont {Menotti}},
  \bibinfo {author} {\bibfnamefont {B.}~\bibnamefont {Capogrosso-Sansone}}, \
  and\ \bibinfo {author} {\bibfnamefont {M.}~\bibnamefont {Lewenstein}},\
  }\href {http://stacks.iop.org/0953-4075/44/i=19/a=193001} {\bibfield
  {journal} {\bibinfo  {journal} {J. Phys. B}\ }\textbf {\bibinfo {volume}
  {44}},\ \bibinfo {pages} {193001} (\bibinfo {year} {2011})}\BibitemShut
  {NoStop}%
\bibitem [{\citenamefont {Baranov}\ \emph {et~al.}(2012)\citenamefont
  {Baranov}, \citenamefont {Dalmonte}, \citenamefont {Pupillo},\ and\
  \citenamefont {Zoller}}]{Baranov12}%
  \BibitemOpen
  \bibfield  {author} {\bibinfo {author} {\bibfnamefont {M.~A.}\ \bibnamefont
  {Baranov}}, \bibinfo {author} {\bibfnamefont {M.}~\bibnamefont {Dalmonte}},
  \bibinfo {author} {\bibfnamefont {G.}~\bibnamefont {Pupillo}}, \ and\
  \bibinfo {author} {\bibfnamefont {P.}~\bibnamefont {Zoller}},\ }\href
  {http://pubs.acs.org/doi/abs/10.1021/cr2003568} {\bibfield  {journal}
  {\bibinfo  {journal} {Chem. Rev.}\ }\textbf {\bibinfo {volume} {112}},\
  \bibinfo {pages} {5012} (\bibinfo {year} {2012})}\BibitemShut {NoStop}%
\bibitem [{\citenamefont {K\"{o}hler}\ \emph {et~al.}(2006)\citenamefont
  {K\"{o}hler}, \citenamefont {G\'{o}ral},\ and\ \citenamefont
  {Julienne}}]{Kohler2006}%
  \BibitemOpen
  \bibfield  {author} {\bibinfo {author} {\bibfnamefont {T.}~\bibnamefont
  {K\"{o}hler}}, \bibinfo {author} {\bibfnamefont {K.}~\bibnamefont
  {G\'{o}ral}}, \ and\ \bibinfo {author} {\bibfnamefont {P.~S.}\ \bibnamefont
  {Julienne}},\ }\href@noop {} {\bibfield  {journal} {\bibinfo  {journal} {Rev.
  Mod. Phys.}\ }\textbf {\bibinfo {volume} {78}},\ \bibinfo {eid} {1311}
  (\bibinfo {year} {2006})}\BibitemShut {NoStop}%
\bibitem [{\citenamefont {Chin}\ \emph {et~al.}(2010)\citenamefont {Chin},
  \citenamefont {Grimm}, \citenamefont {Julienne},\ and\ \citenamefont
  {Tiesinga}}]{Chin10}%
  \BibitemOpen
  \bibfield  {author} {\bibinfo {author} {\bibfnamefont {C.}~\bibnamefont
  {Chin}}, \bibinfo {author} {\bibfnamefont {R.}~\bibnamefont {Grimm}},
  \bibinfo {author} {\bibfnamefont {P.}~\bibnamefont {Julienne}}, \ and\
  \bibinfo {author} {\bibfnamefont {E.}~\bibnamefont {Tiesinga}},\ }\href
  {\doibase 10.1103/RevModPhys.82.1225} {\bibfield  {journal} {\bibinfo
  {journal} {Rev. Mod. Phys.}\ }\textbf {\bibinfo {volume} {82}},\ \bibinfo
  {pages} {1225} (\bibinfo {year} {2010})}\BibitemShut {NoStop}%
\bibitem [{\citenamefont {Bergmann}\ \emph {et~al.}(1998)\citenamefont
  {Bergmann}, \citenamefont {Theuer},\ and\ \citenamefont {Shore}}]{Bergman98}%
  \BibitemOpen
  \bibfield  {author} {\bibinfo {author} {\bibfnamefont {K.}~\bibnamefont
  {Bergmann}}, \bibinfo {author} {\bibfnamefont {H.}~\bibnamefont {Theuer}}, \
  and\ \bibinfo {author} {\bibfnamefont {B.~W.}\ \bibnamefont {Shore}},\
  }\href@noop {} {\bibfield  {journal} {\bibinfo  {journal} {Rev. Mod. Phys.}\
  }\textbf {\bibinfo {volume} {70}},\ \bibinfo {pages} {1003} (\bibinfo {year}
  {1998})}\BibitemShut {NoStop}%
\bibitem [{\citenamefont {Ni}\ \emph {et~al.}(2008)\citenamefont {Ni},
  \citenamefont {Ospelkaus}, \citenamefont {de~Miranda}, \citenamefont {Pe'er},
  \citenamefont {Neyenhuis}, \citenamefont {Zirbel}, \citenamefont
  {Kotochigova}, \citenamefont {Julienne}, \citenamefont {Jin},\ and\
  \citenamefont {Ye}}]{Ni2008}%
  \BibitemOpen
  \bibfield  {author} {\bibinfo {author} {\bibfnamefont {K.-K.}\ \bibnamefont
  {Ni}}, \bibinfo {author} {\bibfnamefont {S.}~\bibnamefont {Ospelkaus}},
  \bibinfo {author} {\bibfnamefont {M.~H.~G.}\ \bibnamefont {de~Miranda}},
  \bibinfo {author} {\bibfnamefont {A.}~\bibnamefont {Pe'er}}, \bibinfo
  {author} {\bibfnamefont {B.}~\bibnamefont {Neyenhuis}}, \bibinfo {author}
  {\bibfnamefont {J.~J.}\ \bibnamefont {Zirbel}}, \bibinfo {author}
  {\bibfnamefont {S.}~\bibnamefont {Kotochigova}}, \bibinfo {author}
  {\bibfnamefont {P.~S.}\ \bibnamefont {Julienne}}, \bibinfo {author}
  {\bibfnamefont {D.~S.}\ \bibnamefont {Jin}}, \ and\ \bibinfo {author}
  {\bibfnamefont {J.}~\bibnamefont {Ye}},\ }\href {\doibase
  10.1126/science.1163861} {\bibfield  {journal} {\bibinfo  {journal}
  {Science}\ }\textbf {\bibinfo {volume} {322}},\ \bibinfo {pages} {231}
  (\bibinfo {year} {2008})}\BibitemShut {NoStop}%
\bibitem [{\citenamefont {Park}\ \emph {et~al.}(2015)\citenamefont {Park},
  \citenamefont {Will},\ and\ \citenamefont {Zwierlein}}]{Park2015}%
  \BibitemOpen
  \bibfield  {author} {\bibinfo {author} {\bibfnamefont {J.~W.}\ \bibnamefont
  {Park}}, \bibinfo {author} {\bibfnamefont {S.~A.}\ \bibnamefont {Will}}, \
  and\ \bibinfo {author} {\bibfnamefont {M.~W.}\ \bibnamefont {Zwierlein}},\
  }\href {\doibase 10.1103/PhysRevLett.114.205302} {\bibfield  {journal}
  {\bibinfo  {journal} {Phys. Rev. Lett.}\ }\textbf {\bibinfo {volume} {114}},\
  \bibinfo {pages} {205302} (\bibinfo {year} {2015})}\BibitemShut {NoStop}%
\bibitem [{\citenamefont {Takekoshi}\ \emph {et~al.}(2014)\citenamefont
  {Takekoshi}, \citenamefont {Reichs\"ollner}, \citenamefont {Schindewolf},
  \citenamefont {Hutson}, \citenamefont {Le~Sueur}, \citenamefont {Dulieu},
  \citenamefont {Ferlaino}, \citenamefont {Grimm},\ and\ \citenamefont
  {N\"agerl}}]{Takekoshi2014}%
  \BibitemOpen
  \bibfield  {author} {\bibinfo {author} {\bibfnamefont {T.}~\bibnamefont
  {Takekoshi}}, \bibinfo {author} {\bibfnamefont {L.}~\bibnamefont
  {Reichs\"ollner}}, \bibinfo {author} {\bibfnamefont {A.}~\bibnamefont
  {Schindewolf}}, \bibinfo {author} {\bibfnamefont {J.~M.}\ \bibnamefont
  {Hutson}}, \bibinfo {author} {\bibfnamefont {C.~R.}\ \bibnamefont
  {Le~Sueur}}, \bibinfo {author} {\bibfnamefont {O.}~\bibnamefont {Dulieu}},
  \bibinfo {author} {\bibfnamefont {F.}~\bibnamefont {Ferlaino}}, \bibinfo
  {author} {\bibfnamefont {R.}~\bibnamefont {Grimm}}, \ and\ \bibinfo {author}
  {\bibfnamefont {H.-C.}\ \bibnamefont {N\"agerl}},\ }\href {\doibase
  10.1103/PhysRevLett.113.205301} {\bibfield  {journal} {\bibinfo  {journal}
  {Phys. Rev. Lett.}\ }\textbf {\bibinfo {volume} {113}},\ \bibinfo {pages}
  {205301} (\bibinfo {year} {2014})}\BibitemShut {NoStop}%
\bibitem [{\citenamefont {Molony}\ \emph {et~al.}(2014)\citenamefont {Molony},
  \citenamefont {Gregory}, \citenamefont {Ji}, \citenamefont {Lu},
  \citenamefont {K\"oppinger}, \citenamefont {Le~Sueur}, \citenamefont
  {Blackley}, \citenamefont {Hutson},\ and\ \citenamefont
  {Cornish}}]{Molony2014}%
  \BibitemOpen
  \bibfield  {author} {\bibinfo {author} {\bibfnamefont {P.~K.}\ \bibnamefont
  {Molony}}, \bibinfo {author} {\bibfnamefont {P.~D.}\ \bibnamefont {Gregory}},
  \bibinfo {author} {\bibfnamefont {Z.}~\bibnamefont {Ji}}, \bibinfo {author}
  {\bibfnamefont {B.}~\bibnamefont {Lu}}, \bibinfo {author} {\bibfnamefont
  {M.~P.}\ \bibnamefont {K\"oppinger}}, \bibinfo {author} {\bibfnamefont
  {C.~R.}\ \bibnamefont {Le~Sueur}}, \bibinfo {author} {\bibfnamefont {C.~L.}\
  \bibnamefont {Blackley}}, \bibinfo {author} {\bibfnamefont {J.~M.}\
  \bibnamefont {Hutson}}, \ and\ \bibinfo {author} {\bibfnamefont {S.~L.}\
  \bibnamefont {Cornish}},\ }\href {\doibase 10.1103/PhysRevLett.113.255301}
  {\bibfield  {journal} {\bibinfo  {journal} {Phys. Rev. Lett.}\ }\textbf
  {\bibinfo {volume} {113}},\ \bibinfo {pages} {255301} (\bibinfo {year}
  {2014})}\BibitemShut {NoStop}%
\bibitem [{\citenamefont {\ifmmode~\dot{Z}\else \.{Z}\fi{}uchowski}\ and\
  \citenamefont {Hutson}(2010)}]{Zuchowski2010}%
  \BibitemOpen
  \bibfield  {author} {\bibinfo {author} {\bibfnamefont {P.~S.}\ \bibnamefont
  {\ifmmode~\dot{Z}\else \.{Z}\fi{}uchowski}}\ and\ \bibinfo {author}
  {\bibfnamefont {J.~M.}\ \bibnamefont {Hutson}},\ }\href {\doibase
  10.1103/PhysRevA.81.060703} {\bibfield  {journal} {\bibinfo  {journal} {Phys.
  Rev. A}\ }\textbf {\bibinfo {volume} {81}},\ \bibinfo {pages} {060703}
  (\bibinfo {year} {2010})}\BibitemShut {NoStop}%
\bibitem [{\citenamefont {Ospelkaus}\ \emph {et~al.}(2010)\citenamefont
  {Ospelkaus}, \citenamefont {Ni}, \citenamefont {Wang}, \citenamefont
  {de~Miranda}, \citenamefont {Neyenhuis}, \citenamefont {Qu{\'e}m{\'e}ner},
  \citenamefont {Julienne}, \citenamefont {Bohn}, \citenamefont {Jin},\ and\
  \citenamefont {Ye}}]{Ospelkaus10}%
  \BibitemOpen
  \bibfield  {author} {\bibinfo {author} {\bibfnamefont {S.}~\bibnamefont
  {Ospelkaus}}, \bibinfo {author} {\bibfnamefont {K.-K.}\ \bibnamefont {Ni}},
  \bibinfo {author} {\bibfnamefont {D.}~\bibnamefont {Wang}}, \bibinfo {author}
  {\bibfnamefont {M.~H.~G.}\ \bibnamefont {de~Miranda}}, \bibinfo {author}
  {\bibfnamefont {B.}~\bibnamefont {Neyenhuis}}, \bibinfo {author}
  {\bibfnamefont {G.}~\bibnamefont {Qu{\'e}m{\'e}ner}}, \bibinfo {author}
  {\bibfnamefont {P.~S.}\ \bibnamefont {Julienne}}, \bibinfo {author}
  {\bibfnamefont {J.~L.}\ \bibnamefont {Bohn}}, \bibinfo {author}
  {\bibfnamefont {D.~S.}\ \bibnamefont {Jin}}, \ and\ \bibinfo {author}
  {\bibfnamefont {J.}~\bibnamefont {Ye}},\ }\href@noop {} {\bibfield  {journal}
  {\bibinfo  {journal} {Science}\ }\textbf {\bibinfo {volume} {327}},\ \bibinfo
  {pages} {853} (\bibinfo {year} {2010})}\BibitemShut {NoStop}%
\bibitem [{\citenamefont {Ni}\ \emph {et~al.}(2010)\citenamefont {Ni},
  \citenamefont {Ospelkaus}, \citenamefont {Wang}, \citenamefont
  {Qu{\'e}m{\'e}ner}, \citenamefont {Neyenhuis}, \citenamefont {de~Miranda},
  \citenamefont {Bohn}, \citenamefont {Ye},\ and\ \citenamefont {Jin}}]{Ni10}%
  \BibitemOpen
  \bibfield  {author} {\bibinfo {author} {\bibfnamefont {K.-K.}\ \bibnamefont
  {Ni}}, \bibinfo {author} {\bibfnamefont {S.}~\bibnamefont {Ospelkaus}},
  \bibinfo {author} {\bibfnamefont {D.}~\bibnamefont {Wang}}, \bibinfo {author}
  {\bibfnamefont {G.}~\bibnamefont {Qu{\'e}m{\'e}ner}}, \bibinfo {author}
  {\bibfnamefont {B.}~\bibnamefont {Neyenhuis}}, \bibinfo {author}
  {\bibfnamefont {M.~H.~G.}\ \bibnamefont {de~Miranda}}, \bibinfo {author}
  {\bibfnamefont {J.~L.}\ \bibnamefont {Bohn}}, \bibinfo {author}
  {\bibfnamefont {J.}~\bibnamefont {Ye}}, \ and\ \bibinfo {author}
  {\bibfnamefont {D.~S.}\ \bibnamefont {Jin}},\ }\href@noop {} {\bibfield
  {journal} {\bibinfo  {journal} {Nature}\ }\textbf {\bibinfo {volume} {464}},\
  \bibinfo {pages} {1324} (\bibinfo {year} {2010})}\BibitemShut {NoStop}%
\bibitem [{\citenamefont {Mayle}\ \emph {et~al.}(2013)\citenamefont {Mayle},
  \citenamefont {Qu\'em\'ener}, \citenamefont {Ruzic},\ and\ \citenamefont
  {Bohn}}]{Mayle2013}%
  \BibitemOpen
  \bibfield  {author} {\bibinfo {author} {\bibfnamefont {M.}~\bibnamefont
  {Mayle}}, \bibinfo {author} {\bibfnamefont {G.}~\bibnamefont {Qu\'em\'ener}},
  \bibinfo {author} {\bibfnamefont {B.~P.}\ \bibnamefont {Ruzic}}, \ and\
  \bibinfo {author} {\bibfnamefont {J.~L.}\ \bibnamefont {Bohn}},\ }\href
  {\doibase 10.1103/PhysRevA.87.012709} {\bibfield  {journal} {\bibinfo
  {journal} {Phys. Rev. A}\ }\textbf {\bibinfo {volume} {87}},\ \bibinfo
  {pages} {012709} (\bibinfo {year} {2013})}\BibitemShut {NoStop}%
\bibitem [{\citenamefont {Pashov}\ \emph {et~al.}(2005)\citenamefont {Pashov},
  \citenamefont {Docenko}, \citenamefont {Tamanis}, \citenamefont {Ferber},
  \citenamefont {Kn\"ockel},\ and\ \citenamefont {Tiemann}}]{Pashov2005}%
  \BibitemOpen
  \bibfield  {author} {\bibinfo {author} {\bibfnamefont {A.}~\bibnamefont
  {Pashov}}, \bibinfo {author} {\bibfnamefont {O.}~\bibnamefont {Docenko}},
  \bibinfo {author} {\bibfnamefont {M.}~\bibnamefont {Tamanis}}, \bibinfo
  {author} {\bibfnamefont {R.}~\bibnamefont {Ferber}}, \bibinfo {author}
  {\bibfnamefont {H.}~\bibnamefont {Kn\"ockel}}, \ and\ \bibinfo {author}
  {\bibfnamefont {E.}~\bibnamefont {Tiemann}},\ }\href {\doibase
  10.1103/PhysRevA.72.062505} {\bibfield  {journal} {\bibinfo  {journal} {Phys.
  Rev. A}\ }\textbf {\bibinfo {volume} {72}},\ \bibinfo {pages} {062505}
  (\bibinfo {year} {2005})}\BibitemShut {NoStop}%
\bibitem [{\citenamefont {Seto}\ \emph {et~al.}(2000)\citenamefont {Seto},
  \citenamefont {Le~Roy}, \citenamefont {Vergès},\ and\ \citenamefont
  {Amiot}}]{Seto2000}%
  \BibitemOpen
  \bibfield  {author} {\bibinfo {author} {\bibfnamefont {J.~Y.}\ \bibnamefont
  {Seto}}, \bibinfo {author} {\bibfnamefont {R.~J.}\ \bibnamefont {Le~Roy}},
  \bibinfo {author} {\bibfnamefont {J.}~\bibnamefont {Vergès}}, \ and\ \bibinfo
  {author} {\bibfnamefont {C.}~\bibnamefont {Amiot}},\ }\href {\doibase
  http://dx.doi.org/10.1063/1.1286979} {\bibfield  {journal} {\bibinfo
  {journal} {J. Chem. Phys.}\ }\textbf {\bibinfo {volume} {113}},\ \bibinfo
  {pages} {3067} (\bibinfo {year} {2000})}\BibitemShut {NoStop}%
\bibitem [{\citenamefont {Jones}\ \emph {et~al.}(1996)\citenamefont {Jones},
  \citenamefont {Maleki}, \citenamefont {Bize}, \citenamefont {Lett},
  \citenamefont {Williams}, \citenamefont {Richling}, \citenamefont
  {Kn\"ockel}, \citenamefont {Tiemann}, \citenamefont {Wang}, \citenamefont
  {Gould},\ and\ \citenamefont {Stwalley}}]{Jones96}%
  \BibitemOpen
  \bibfield  {author} {\bibinfo {author} {\bibfnamefont {K.~M.}\ \bibnamefont
  {Jones}}, \bibinfo {author} {\bibfnamefont {S.}~\bibnamefont {Maleki}},
  \bibinfo {author} {\bibfnamefont {S.}~\bibnamefont {Bize}}, \bibinfo {author}
  {\bibfnamefont {P.~D.}\ \bibnamefont {Lett}}, \bibinfo {author}
  {\bibfnamefont {C.~J.}\ \bibnamefont {Williams}}, \bibinfo {author}
  {\bibfnamefont {H.}~\bibnamefont {Richling}}, \bibinfo {author}
  {\bibfnamefont {H.}~\bibnamefont {Kn\"ockel}}, \bibinfo {author}
  {\bibfnamefont {E.}~\bibnamefont {Tiemann}}, \bibinfo {author} {\bibfnamefont
  {H.}~\bibnamefont {Wang}}, \bibinfo {author} {\bibfnamefont {P.~L.}\
  \bibnamefont {Gould}}, \ and\ \bibinfo {author} {\bibfnamefont {W.~C.}\
  \bibnamefont {Stwalley}},\ }\href {\doibase 10.1103/PhysRevA.54.R1006}
  {\bibfield  {journal} {\bibinfo  {journal} {Phys. Rev. A}\ }\textbf {\bibinfo
  {volume} {54}},\ \bibinfo {pages} {R1006} (\bibinfo {year}
  {1996})}\BibitemShut {NoStop}%
\bibitem [{not()}]{note0}%
  \BibitemOpen
  \href@noop {} {}\bibinfo {note} {However, for two $^{23}$Na$^{87}$Rb molecules in the first excited vibrational level $v''=1$,
	the same reaction is already exothermic~\cite{Pashov2005,Seto2000,Jones96}. It is thus possible to investigate both reactive and non-reactive	behavior in the same experimental settings. This feature also exists in the other two recently created chemically stable alkali polar molecules $^{87}$Rb$^{133}$Cs and $^{23}$Na$^{40}$K.}\BibitemShut {Stop}%
\bibitem [{\citenamefont {Dagdigian}\ and\ \citenamefont
  {Wharton}(1972)}]{Dagdigian1972}%
  \BibitemOpen
  \bibfield  {author} {\bibinfo {author} {\bibfnamefont {P.~J.}\ \bibnamefont
  {Dagdigian}}\ and\ \bibinfo {author} {\bibfnamefont {L.}~\bibnamefont
  {Wharton}},\ }\href {\doibase http://dx.doi.org/10.1063/1.1678429} {\bibfield
   {journal} {\bibinfo  {journal} {J. Chem. Phys.}\ }\textbf {\bibinfo {volume}
  {57}},\ \bibinfo {pages} {1487} (\bibinfo {year} {1972})}\BibitemShut
  {NoStop}%
\bibitem [{\citenamefont {Igel-Mann}\ \emph {et~al.}(1986)\citenamefont
  {Igel-Mann}, \citenamefont {Wedig}, \citenamefont {Fuentealba},\ and\
  \citenamefont {Stoll}}]{IgelMann1986}%
  \BibitemOpen
  \bibfield  {author} {\bibinfo {author} {\bibfnamefont {G.}~\bibnamefont
  {Igel-Mann}}, \bibinfo {author} {\bibfnamefont {U.}~\bibnamefont {Wedig}},
  \bibinfo {author} {\bibfnamefont {P.}~\bibnamefont {Fuentealba}}, \ and\
  \bibinfo {author} {\bibfnamefont {H.}~\bibnamefont {Stoll}},\ }\href
  {\doibase http://dx.doi.org/10.1063/1.450649} {\bibfield  {journal} {\bibinfo
   {journal} {J. Chem. Phys.}\ }\textbf {\bibinfo {volume} {84}},\ \bibinfo
  {pages} {5007} (\bibinfo {year} {1986})}\BibitemShut {NoStop}%
\bibitem [{\citenamefont {Tarnovsky}\ \emph {et~al.}(1993)\citenamefont
  {Tarnovsky}, \citenamefont {Bunimovicz}, \citenamefont {Vuskovic},
  \citenamefont {Stumpf},\ and\ \citenamefont {Bederson}}]{Tarnovsky1993}%
  \BibitemOpen
  \bibfield  {author} {\bibinfo {author} {\bibfnamefont {V.}~\bibnamefont
  {Tarnovsky}}, \bibinfo {author} {\bibfnamefont {M.}~\bibnamefont
  {Bunimovicz}}, \bibinfo {author} {\bibfnamefont {L.}~\bibnamefont
  {Vuskovic}}, \bibinfo {author} {\bibfnamefont {B.}~\bibnamefont {Stumpf}}, \
  and\ \bibinfo {author} {\bibfnamefont {B.}~\bibnamefont {Bederson}},\ }\href
  {\doibase http://dx.doi.org/10.1063/1.464017} {\bibfield  {journal} {\bibinfo
   {journal} {J. Chem. Phys.}\ }\textbf {\bibinfo {volume} {98}},\ \bibinfo
  {pages} {3894} (\bibinfo {year} {1993})}\BibitemShut {NoStop}%
\bibitem [{\citenamefont {Aymar}\ and\ \citenamefont
  {Dulieu}(2005)}]{Aymar2005}%
  \BibitemOpen
  \bibfield  {author} {\bibinfo {author} {\bibfnamefont {M.}~\bibnamefont
  {Aymar}}\ and\ \bibinfo {author} {\bibfnamefont {O.}~\bibnamefont {Dulieu}},\
  }\href {http://link.aip.org/link/?JCP/122/204302/1} {\bibfield  {journal}
  {\bibinfo  {journal} {J. Chem. Phys.}\ }\textbf {\bibinfo {volume} {122}},\
  \bibinfo {eid} {204302} (\bibinfo {year} {2005})}\BibitemShut {NoStop}%
\bibitem [{\citenamefont {Gao}(2008)}]{Gao2008}%
  \BibitemOpen
  \bibfield  {author} {\bibinfo {author} {\bibfnamefont {B.}~\bibnamefont
  {Gao}},\ }\href {\doibase 10.1103/PhysRevA.78.012702} {\bibfield  {journal}
  {\bibinfo  {journal} {Phys. Rev. A}\ }\textbf {\bibinfo {volume} {78}},\
  \bibinfo {pages} {012702} (\bibinfo {year} {2008})}\BibitemShut {NoStop}%
\bibitem [{\citenamefont {Wang}\ \emph {et~al.}(2015)\citenamefont {Wang},
  \citenamefont {He}, \citenamefont {Li}, \citenamefont {Zhu}, \citenamefont
  {Chen},\ and\ \citenamefont {Wang}}]{Wangfudong2015}%
  \BibitemOpen
  \bibfield  {author} {\bibinfo {author} {\bibfnamefont {F.}~\bibnamefont
  {Wang}}, \bibinfo {author} {\bibfnamefont {X.}~\bibnamefont {He}}, \bibinfo
  {author} {\bibfnamefont {X.}~\bibnamefont {Li}}, \bibinfo {author}
  {\bibfnamefont {B.}~\bibnamefont {Zhu}}, \bibinfo {author} {\bibfnamefont
  {J.}~\bibnamefont {Chen}}, \ and\ \bibinfo {author} {\bibfnamefont
  {D.}~\bibnamefont {Wang}},\ }\href
  {http://stacks.iop.org/1367-2630/17/i=3/a=035003} {\bibfield  {journal}
  {\bibinfo  {journal} {New J. Phys.}\ }\textbf {\bibinfo {volume} {17}},\
  \bibinfo {pages} {035003} (\bibinfo {year} {2015})}\BibitemShut {NoStop}%
\bibitem [{\citenamefont {Wang}\ \emph {et~al.}(2013)\citenamefont {Wang},
  \citenamefont {Xiong}, \citenamefont {Li}, \citenamefont {Wang},\ and\
  \citenamefont {Tiemann}}]{Wangfudong2013}%
  \BibitemOpen
  \bibfield  {author} {\bibinfo {author} {\bibfnamefont {F.}~\bibnamefont
  {Wang}}, \bibinfo {author} {\bibfnamefont {D.}~\bibnamefont {Xiong}},
  \bibinfo {author} {\bibfnamefont {X.}~\bibnamefont {Li}}, \bibinfo {author}
  {\bibfnamefont {D.}~\bibnamefont {Wang}}, \ and\ \bibinfo {author}
  {\bibfnamefont {E.}~\bibnamefont {Tiemann}},\ }\href {\doibase
  10.1103/PhysRevA.87.050702} {\bibfield  {journal} {\bibinfo  {journal} {Phys.
  Rev. A(R)}\ }\textbf {\bibinfo {volume} {87}},\ \bibinfo {pages} {050702}
  (\bibinfo {year} {2013})}\BibitemShut {NoStop}%
\bibitem [{\citenamefont {Docenko}\ \emph {et~al.}(2007)\citenamefont
  {Docenko}, \citenamefont {Tamanis}, \citenamefont {Ferber}, \citenamefont
  {Pazyuk}, \citenamefont {Zaitsevskii}, \citenamefont {Stolyarov},
  \citenamefont {Pashov}, \citenamefont {Kn\"ockel},\ and\ \citenamefont
  {Tiemann}}]{Docenko2007}%
  \BibitemOpen
  \bibfield  {author} {\bibinfo {author} {\bibfnamefont {O.}~\bibnamefont
  {Docenko}}, \bibinfo {author} {\bibfnamefont {M.}~\bibnamefont {Tamanis}},
  \bibinfo {author} {\bibfnamefont {R.}~\bibnamefont {Ferber}}, \bibinfo
  {author} {\bibfnamefont {E.~A.}\ \bibnamefont {Pazyuk}}, \bibinfo {author}
  {\bibfnamefont {A.}~\bibnamefont {Zaitsevskii}}, \bibinfo {author}
  {\bibfnamefont {A.~V.}\ \bibnamefont {Stolyarov}}, \bibinfo {author}
  {\bibfnamefont {A.}~\bibnamefont {Pashov}}, \bibinfo {author} {\bibfnamefont
  {H.}~\bibnamefont {Kn\"ockel}}, \ and\ \bibinfo {author} {\bibfnamefont
  {E.}~\bibnamefont {Tiemann}},\ }\href {\doibase 10.1103/PhysRevA.75.042503}
  {\bibfield  {journal} {\bibinfo  {journal} {Phys. Rev. A}\ }\textbf {\bibinfo
  {volume} {75}},\ \bibinfo {pages} {042503} (\bibinfo {year}
  {2007})}\BibitemShut {NoStop}%
\bibitem [{\citenamefont {Drever}\ \emph {et~al.}(1983)\citenamefont {Drever},
  \citenamefont {Hall}, \citenamefont {Kowalski}, \citenamefont {Hough},
  \citenamefont {Ford}, \citenamefont {Munley},\ and\ \citenamefont
  {Ward}}]{Drever1983}%
  \BibitemOpen
  \bibfield  {author} {\bibinfo {author} {\bibfnamefont {R.~W.~P.}\
  \bibnamefont {Drever}}, \bibinfo {author} {\bibfnamefont {J.~L.}\
  \bibnamefont {Hall}}, \bibinfo {author} {\bibfnamefont {F.~V.}\ \bibnamefont
  {Kowalski}}, \bibinfo {author} {\bibfnamefont {J.}~\bibnamefont {Hough}},
  \bibinfo {author} {\bibfnamefont {G.~M.}\ \bibnamefont {Ford}}, \bibinfo
  {author} {\bibfnamefont {A.~J.}\ \bibnamefont {Munley}}, \ and\ \bibinfo
  {author} {\bibfnamefont {H.}~\bibnamefont {Ward}},\ }\href {\doibase
  10.1007/BF00702605} {\bibfield  {journal} {\bibinfo  {journal} {Appl. Phys.
  B}\ }\textbf {\bibinfo {volume} {31}},\ \bibinfo {pages} {97} (\bibinfo
  {year} {1983})}\BibitemShut {NoStop}%
\bibitem [{\citenamefont {Kasahara}\ \emph {et~al.}(1996)\citenamefont
  {Kasahara}, \citenamefont {Ebi}, \citenamefont {Tanimura}, \citenamefont
  {Ikoma}, \citenamefont {Matsubara}, \citenamefont {Baba},\ and\ \citenamefont
  {Kato}}]{Kasahara1996}%
  \BibitemOpen
  \bibfield  {author} {\bibinfo {author} {\bibfnamefont {S.}~\bibnamefont
  {Kasahara}}, \bibinfo {author} {\bibfnamefont {T.}~\bibnamefont {Ebi}},
  \bibinfo {author} {\bibfnamefont {M.}~\bibnamefont {Tanimura}}, \bibinfo
  {author} {\bibfnamefont {H.}~\bibnamefont {Ikoma}}, \bibinfo {author}
  {\bibfnamefont {K.}~\bibnamefont {Matsubara}}, \bibinfo {author}
  {\bibfnamefont {M.}~\bibnamefont {Baba}}, \ and\ \bibinfo {author}
  {\bibfnamefont {H.}~\bibnamefont {Kato}},\ }\href {\doibase
  http://dx.doi.org/10.1063/1.472000} {\bibfield  {journal} {\bibinfo
  {journal} {J. Chem. Phys.}\ }\textbf {\bibinfo {volume} {105}},\ \bibinfo
  {pages} {1341} (\bibinfo {year} {1996})}\BibitemShut {NoStop}%
\bibitem [{Not()}]{note1}%
  \BibitemOpen
  \href@noop {} {}\bibinfo {note} {Detailed discussions on the one-photon and
  two-photon spectroscopy, including deperturbation, hyperfine structure, and
  transition strength calibrations, will be published in another
  paper which is in preparation.}\BibitemShut {Stop}%
\bibitem [{\citenamefont {Zhu}\ \emph {et~al.}(2016)\citenamefont {Zhu},
  \citenamefont {Li}, \citenamefont {He}, \citenamefont {Guo}, \citenamefont
  {Wang}, \citenamefont {Vexiau}, \citenamefont {Bouloufa-Maafa}, \citenamefont
  {Dulieu},\ and\ \citenamefont {Wang}}]{Zhu2016}%
  \BibitemOpen
  \bibfield  {author} {\bibinfo {author} {\bibfnamefont {B.}~\bibnamefont
  {Zhu}}, \bibinfo {author} {\bibfnamefont {X.}~\bibnamefont {Li}}, \bibinfo
  {author} {\bibfnamefont {X.}~\bibnamefont {He}}, \bibinfo {author}
  {\bibfnamefont {M.}~\bibnamefont {Guo}}, \bibinfo {author} {\bibfnamefont
  {F.}~\bibnamefont {Wang}}, \bibinfo {author} {\bibfnamefont {R.}~\bibnamefont
  {Vexiau}}, \bibinfo {author} {\bibfnamefont {N.}~\bibnamefont
  {Bouloufa-Maafa}}, \bibinfo {author} {\bibfnamefont {O.}~\bibnamefont
  {Dulieu}}, \ and\ \bibinfo {author} {\bibfnamefont {D.}~\bibnamefont
  {Wang}},\ }\href {\doibase 10.1103/PhysRevA.93.012508} {\bibfield  {journal}
  {\bibinfo  {journal} {Phys. Rev. A}\ }\textbf {\bibinfo {volume} {93}},\
  \bibinfo {pages} {012508} (\bibinfo {year} {2016})}\BibitemShut {NoStop}%
\bibitem [{\citenamefont {Aldegunde}\ and\ \citenamefont
  {Hutson}(2009)}]{Aldegunde2009}%
  \BibitemOpen
  \bibfield  {author} {\bibinfo {author} {\bibfnamefont {J.}~\bibnamefont
  {Aldegunde}}\ and\ \bibinfo {author} {\bibfnamefont {J.~M.}\ \bibnamefont
  {Hutson}},\ }\href {\doibase 10.1103/PhysRevA.79.013401} {\bibfield
  {journal} {\bibinfo  {journal} {Phys. Rev. A}\ }\textbf {\bibinfo {volume}
  {79}},\ \bibinfo {pages} {013401} (\bibinfo {year} {2009})}\BibitemShut
  {NoStop}%
\bibitem [{not()}]{note3}%
  \BibitemOpen
  \href@noop {} {}\bibinfo {note} {J. Aldegunde and J.~M. Hutson, private communication.}\BibitemShut {Stop}%
\bibitem [{not()}]{note2}%
  \BibitemOpen
  \href@noop {} {}\bibinfo {note} {The sloshing and breathing motions are
  caused by technical imperfections in the overall trap potential, as well as
  the trap potential differences between the Feshbach molecules and the
  ground-state molecules. The trap frequencies are extracted from the sloshing
  motions. After 50 ms, motions along the tightly confined y and z directions
  are damped out, while some residual motions along the x direction persist for
  the whole measurement.}\BibitemShut {Stop}%
\bibitem [{\citenamefont {Lepers}\ \emph {et~al.}(2013)\citenamefont {Lepers},
  \citenamefont {Vexiau}, \citenamefont {Aymar}, \citenamefont
  {Bouloufa-Maafa},\ and\ \citenamefont {Dulieu}}]{Lepers2013}%
  \BibitemOpen
  \bibfield  {author} {\bibinfo {author} {\bibfnamefont {M.}~\bibnamefont
  {Lepers}}, \bibinfo {author} {\bibfnamefont {R.}~\bibnamefont {Vexiau}},
  \bibinfo {author} {\bibfnamefont {M.}~\bibnamefont {Aymar}}, \bibinfo
  {author} {\bibfnamefont {N.}~\bibnamefont {Bouloufa-Maafa}}, \ and\ \bibinfo
  {author} {\bibfnamefont {O.}~\bibnamefont {Dulieu}},\ }\href {\doibase
  10.1103/PhysRevA.88.032709} {\bibfield  {journal} {\bibinfo  {journal} {Phys.
  Rev. A}\ }\textbf {\bibinfo {volume} {88}},\ \bibinfo {pages} {032709}
  (\bibinfo {year} {2013})}\BibitemShut {NoStop}%
\bibitem [{\citenamefont {Wang}\ and\ \citenamefont
  {Qu\'em\'ener}(2015)}]{Wanggaoren2015}%
  \BibitemOpen
  \bibfield  {author} {\bibinfo {author} {\bibfnamefont {G.}~\bibnamefont
  {Wang}}\ and\ \bibinfo {author} {\bibfnamefont {G.}~\bibnamefont
  {Qu\'em\'ener}},\ }\href {http://stacks.iop.org/1367-2630/17/i=3/a=035015}
  {\bibfield  {journal} {\bibinfo  {journal} {New J. Phys.}\ }\textbf {\bibinfo
  {volume} {17}},\ \bibinfo {pages} {035015} (\bibinfo {year}
  {2015})}\BibitemShut {NoStop}%
\bibitem [{\citenamefont {Avdeenkov}\ \emph {et~al.}(2006)\citenamefont
  {Avdeenkov}, \citenamefont {Kajita},\ and\ \citenamefont
  {Bohn}}]{Avdeenkov2006}%
  \BibitemOpen
  \bibfield  {author} {\bibinfo {author} {\bibfnamefont {A.~V.}\ \bibnamefont
  {Avdeenkov}}, \bibinfo {author} {\bibfnamefont {M.}~\bibnamefont {Kajita}}, \
  and\ \bibinfo {author} {\bibfnamefont {J.~L.}\ \bibnamefont {Bohn}},\ }\href
  {\doibase 10.1103/PhysRevA.73.022707} {\bibfield  {journal} {\bibinfo
  {journal} {Phys. Rev. A}\ }\textbf {\bibinfo {volume} {73}},\ \bibinfo
  {pages} {022707} (\bibinfo {year} {2006})}\BibitemShut {NoStop}%
\bibitem [{\citenamefont {Qu\'em\'ener}\ and\ \citenamefont
  {Bohn}(2016)}]{Quemener2016}%
  \BibitemOpen
  \bibfield  {author} {\bibinfo {author} {\bibfnamefont {G.}~\bibnamefont
  {Qu\'em\'ener}}\ and\ \bibinfo {author} {\bibfnamefont {J.~L.}\ \bibnamefont
  {Bohn}},\ }\href {\doibase 10.1103/PhysRevA.93.012704} {\bibfield  {journal}
  {\bibinfo  {journal} {Phys. Rev. A}\ }\textbf {\bibinfo {volume} {93}},\
  \bibinfo {pages} {012704} (\bibinfo {year} {2016})}\BibitemShut {NoStop}%
\bibitem [{\citenamefont {Danzl}\ \emph {et~al.}(2010)\citenamefont {Danzl},
  \citenamefont {Mark}, \citenamefont {Haller}, \citenamefont {Gustavsson},
  \citenamefont {Hart}, \citenamefont {Aldegunde}, \citenamefont {Hutson},\
  and\ \citenamefont {N\"{a}gerl}}]{Danzl2010}%
  \BibitemOpen
  \bibfield  {author} {\bibinfo {author} {\bibfnamefont {J.}~\bibnamefont
  {Danzl}}, \bibinfo {author} {\bibfnamefont {M.}~\bibnamefont {Mark}},
  \bibinfo {author} {\bibfnamefont {E.}~\bibnamefont {Haller}}, \bibinfo
  {author} {\bibfnamefont {M.}~\bibnamefont {Gustavsson}}, \bibinfo {author}
  {\bibfnamefont {R.}~\bibnamefont {Hart}}, \bibinfo {author} {\bibfnamefont
  {J.}~\bibnamefont {Aldegunde}}, \bibinfo {author} {\bibfnamefont
  {J.}~\bibnamefont {Hutson}}, \ and\ \bibinfo {author} {\bibfnamefont {H.-C.}\
  \bibnamefont {N\"{a}gerl}},\ }\href@noop {} {\bibfield  {journal} {\bibinfo
  {journal} {Nature Phys.}\ }\textbf {\bibinfo {volume} {6}},\ \bibinfo {pages}
  {265} (\bibinfo {year} {2010})}\BibitemShut {NoStop}%
\bibitem [{\citenamefont {Chotia}\ \emph {et~al.}(2012)\citenamefont {Chotia},
  \citenamefont {Neyenhuis}, \citenamefont {Moses}, \citenamefont {Yan},
  \citenamefont {Covey}, \citenamefont {Foss-Feig}, \citenamefont {Rey},
  \citenamefont {Jin},\ and\ \citenamefont {Ye}}]{Chotia2012}%
  \BibitemOpen
  \bibfield  {author} {\bibinfo {author} {\bibfnamefont {A.}~\bibnamefont
  {Chotia}}, \bibinfo {author} {\bibfnamefont {B.}~\bibnamefont {Neyenhuis}},
  \bibinfo {author} {\bibfnamefont {S.~A.}\ \bibnamefont {Moses}}, \bibinfo
  {author} {\bibfnamefont {B.}~\bibnamefont {Yan}}, \bibinfo {author}
  {\bibfnamefont {J.~P.}\ \bibnamefont {Covey}}, \bibinfo {author}
  {\bibfnamefont {M.}~\bibnamefont {Foss-Feig}}, \bibinfo {author}
  {\bibfnamefont {A.~M.}\ \bibnamefont {Rey}}, \bibinfo {author} {\bibfnamefont
  {D.~S.}\ \bibnamefont {Jin}}, \ and\ \bibinfo {author} {\bibfnamefont
  {J.}~\bibnamefont {Ye}},\ }\href {\doibase 10.1103/PhysRevLett.108.080405}
  {\bibfield  {journal} {\bibinfo  {journal} {Phys. Rev. Lett.}\ }\textbf
  {\bibinfo {volume} {108}},\ \bibinfo {pages} {080405} (\bibinfo {year}
  {2012})}\BibitemShut {NoStop}%
\bibitem [{\citenamefont {B\"uchler}\ \emph {et~al.}(2007)\citenamefont
  {B\"uchler}, \citenamefont {Demler}, \citenamefont {Lukin}, \citenamefont
  {Micheli}, \citenamefont {Prokof'ev}, \citenamefont {Pupillo},\ and\
  \citenamefont {Zoller}}]{Buchler2007}%
  \BibitemOpen
  \bibfield  {author} {\bibinfo {author} {\bibfnamefont {H.~P.}\ \bibnamefont
  {B\"uchler}}, \bibinfo {author} {\bibfnamefont {E.}~\bibnamefont {Demler}},
  \bibinfo {author} {\bibfnamefont {M.}~\bibnamefont {Lukin}}, \bibinfo
  {author} {\bibfnamefont {A.}~\bibnamefont {Micheli}}, \bibinfo {author}
  {\bibfnamefont {N.}~\bibnamefont {Prokof'ev}}, \bibinfo {author}
  {\bibfnamefont {G.}~\bibnamefont {Pupillo}}, \ and\ \bibinfo {author}
  {\bibfnamefont {P.}~\bibnamefont {Zoller}},\ }\href {\doibase
  10.1103/PhysRevLett.98.060404} {\bibfield  {journal} {\bibinfo  {journal}
  {Phys. Rev. Lett.}\ }\textbf {\bibinfo {volume} {98}},\ \bibinfo {pages}
  {060404} (\bibinfo {year} {2007})}\BibitemShut {NoStop}%
\bibitem [{\citenamefont {Yi}\ \emph {et~al.}(2007)\citenamefont {Yi},
  \citenamefont {Li},\ and\ \citenamefont {Sun}}]{Yi2007}%
  \BibitemOpen
  \bibfield  {author} {\bibinfo {author} {\bibfnamefont {S.}~\bibnamefont
  {Yi}}, \bibinfo {author} {\bibfnamefont {T.}~\bibnamefont {Li}}, \ and\
  \bibinfo {author} {\bibfnamefont {C.~P.}\ \bibnamefont {Sun}},\ }\href
  {\doibase 10.1103/PhysRevLett.98.260405} {\bibfield  {journal} {\bibinfo
  {journal} {Phys. Rev. Lett.}\ }\textbf {\bibinfo {volume} {98}},\ \bibinfo
  {pages} {260405} (\bibinfo {year} {2007})}\BibitemShut {NoStop}%
\bibitem [{\citenamefont {Capogrosso-Sansone}\ \emph
  {et~al.}(2010)\citenamefont {Capogrosso-Sansone}, \citenamefont {Trefzger},
  \citenamefont {Lewenstein}, \citenamefont {Zoller},\ and\ \citenamefont
  {Pupillo}}]{Capogrosso2010}%
  \BibitemOpen
  \bibfield  {author} {\bibinfo {author} {\bibfnamefont {B.}~\bibnamefont
  {Capogrosso-Sansone}}, \bibinfo {author} {\bibfnamefont {C.}~\bibnamefont
  {Trefzger}}, \bibinfo {author} {\bibfnamefont {M.}~\bibnamefont
  {Lewenstein}}, \bibinfo {author} {\bibfnamefont {P.}~\bibnamefont {Zoller}},
  \ and\ \bibinfo {author} {\bibfnamefont {G.}~\bibnamefont {Pupillo}},\ }\href
  {\doibase 10.1103/PhysRevLett.104.125301} {\bibfield  {journal} {\bibinfo
  {journal} {Phys. Rev. Lett.}\ }\textbf {\bibinfo {volume} {104}},\ \bibinfo
  {pages} {125301} (\bibinfo {year} {2010})}\BibitemShut {NoStop}%
\end{thebibliography}

%

\end{document}